\def\Roman#1{\uppercase\expandafter{\romannumeral#1}}
\documentclass[a4paper,14pt]{article}
\usepackage{amssymb,amsfonts}
\usepackage[mathscr]{eucal}

\title{
On the geometrical representation
of the path integral reduction Jacobian:
The case of dependent coordinates in the description of the reduced motion\\
}

\author{S. N. Storchak\footnote{E-mail adress: storchak@ihep.ru}\\
\small{Institute for High Energy Physics, Protvino, Moscow Region,142284,Russia}}

\begin{document}

\maketitle

\begin{abstract}
The geometrical representation of the path integral reduction Jacobian  
obtained in 
the problem of the path integral quantization of a scalar particle motion 
on a smooth compact Riemannian  manifold with the given free isometric action of the compact semisimple Lie group
has been found for the case when  the local reduced motion 
is described by  means of dependent coordinates.
The result is based on the
 scalar curvature formula    for  
the original manifold which is  viewed as a total space of the principal fibre bundle.

\end{abstract}
{\bf{keywords:}} Marsden-Weinstein reduction, Kaluza--Klein theories, path integral, stochastic analysis.\\

\section{Introduction}

In a well-known approach to the path integral quantization of the gauge theories \cite{Faddeev}, the dynamic of  true degrees of freedom, that are given by the gauge invariant variables, is presented by means of the evolution of the dependent variables defined on a gauge surface. 
The action of the gauge group on the space of the gauge fields, being viewed as a free isometric action of this group on the Hilbert manifold of the gauge fields,  leads to the fibre bundle picture.

The choice of the local coordinates on a total space of the 
principal fibre bundle is performed by 
making use of the gauge conditions by 
which the  local sections of the principal fibre bundle can be defined. 
Since in general  we cannot 
``resolve the gauge'' 
and  determine the local coordinates on a gauge surface  in terms of the explicitly definable functions,  we are forced to use the dependent variables for  description of the true evolution.

In path integrals, the transition to  new variables and the consequent  restriction of the evolution to the gauge surface, being the path integral transformations,\footnote{These  transformations are called  the path integral reduction.}   may not be 
 a quite correct procedure.
The  reason of it is that at present 
there is no a rigorously defined path integral measure on the space of gauge fields.

In a finite-dimensional case, the quantum reduction problem in
 dynamical systems with a symmetry leads to the analogous  path integral transformations.
But in this case, the problem of the path integral reduction can be resolved \cite{Storchak_1,Storchak_2}, since 
we know (at least for the Wiener path integrals) how to define the path integral measure and some of its transformations.
Being established for the finite-dimensional dynamical systems, 
the rules of 
the path integral transformations in 
the reduction procedure 
could be extended to the path integrals used in gauge theories.

With this end in view, we have studied \cite{Storchak_2}
a finite-dimensional dynamical system which is close by its properties to the gauge theories.
This system describes a motion  of a scalar particle on a smooth compact finite-dimensional Riemannian manifold with a  given free isometric action of a compact semisimple Lie group. 

The path integral reduction procedure was realized as the  transformations of the original Wiener path integral, representing the diffusion (or the ``quantum evolution'')  of a scalar particle, to the path integral which determines the  ``quantum evolution'' of a new  dynamical system (a reduced one) given on the orbit space. 
Our path integral transformations result in  the integral relations between both path integrals.
Also, a non-invariance of the path integral measure under the reduction has been found.
The obtained Jacobian gives rise to the additional potential term  
in the Hamilton operator of the reduced dynamical system.

The purpose of the present paper is to find the geometrical representation of the reduction Jacobian. Our derivation of such a representation  will be based on the formula which expresses the scalar curvature of 
the Riemannian  manifold, which is
a total space 
of the principal fibre bundle, 
in terms of the geometrical data characterizing
this bundle.
This formula is similar to the formula for the scalar curvature of the Riemannian manifold with the given Kaluza--Klein metric. 

The paper will be organized as follows. In Section 2,  we  give the basic definitions  and the brief review of the results 
obtained in \cite{Storchak_2}. Besides, in Section 3, by making use of the It\^o's identity, we rewrite  the exponential of the Jacobian 
to replace the stochastic integral of the Jacobian for the ordinary integral taken with respect to the time variable.  

Next section deals with the derivation of the  scalar curvature formula for our Riemannian manifold. 
By transforming the original coordinate basis to the horizontal lift basis,\footnote{In this 
nonholonomic basis  the original metric takes the block-diagonal form.} 
we  calculate the Christoffel coefficients, the Ricci curvature, 
and the scalar curvature.

In  Section 5, by using the scalar curvature formula obtained in the previouse section, we rewrite the reduction Jacobian.

A possible application of the obtained geometrical representation of the Jacobian is  discussed in Conclusion.  

\section{Definitions}
In \cite{Storchak_2}, the diffusion of a scalar particle  on a smooth compact Riemannian manifold $\cal P$ 
has been  considered.  In case of a given free  isometric smooth action of a semisimple compact Lie group $\cal G$ on this manifold we can regard  the manifold $\cal P$ as a total space of the principal fibre bundle $\pi : \cal P \to {\cal P}/{\cal G}=\cal M$. It means that on $\cal P$ 
it is possible to 
introduce new coordinates that are related to the fibre  bundle.

 The original coordinates $Q^A$  given on a local chart of the manifold $\cal P$ have been transformed
for the special coordinates $(Q^{\ast}{}^A,a^{\alpha})$ ($A=1,\ldots , N_{\cal P},N_{\cal P}=\dim {\cal P} ;{\alpha}=1,\ldots , N_{\cal G},N_{\cal G}=\dim {\cal G}$). 
In order for the transformation to be a one-to-one mapping,
the coordinates $Q^{\ast}{}^A$ must be subjected to the additional constraints: ${\chi}^{\alpha}(Q^{\ast})=0$. These constraints are chosen in such a way to define the local submanifolds in the  manifold $\cal P$. We have assumed that these local submanifolds determine the global manifold $\Sigma $. Hence, our principal fibre bundle $P({\cal M},\cal G)$ is trivial. Since it is locally isomorphic to the trivial bundle 
$\Sigma\times {\cal G}\to{\Sigma} $, the coordinates $Q^{\ast}{}^A$ can be used for the coordinatization of the manifold $\cal M$ -- the base of the fibre bundle.
This transformation is fulfilled
at the first step of the reduction procedure.

In new coordinate basis $(\frac{\partial}{\partial Q^{\ast}{}^A},\frac{\partial}{\partial a^{\alpha}})$,  the original metric ${\tilde G}_{\cal A\cal B}(Q^{\ast},a)$ of the manifold $\cal P$ becomes as follows:
\begin{equation}
\left(
\begin{array}{cc}
G_{CD}(Q^{\ast})(P_{\perp})^{C}_{A}
(P_{\perp})^{D}_{B} & G_{CD}(Q^{\ast})(P_{\perp})^
{D}_{A}K^{C}_{\mu}\bar{u}^{\mu}_{\alpha}(a) \\
G_{CD}(Q^{\ast})(P_{\perp})^
{C}_{A}K^{D}_{\nu}\bar{u}^{\nu}_{\beta}(a) & {\gamma }_{\mu \nu }
(Q^{\ast})\bar{u}_\alpha ^\mu (a)\bar{u}_\beta ^\nu (a)
\end{array}
\right),
\label{8}
\end{equation}
where 
$G_{CD}(Q^{\ast})\equiv G_{CD}(F(Q^{\ast},e))$ is given by
\[
G_{CD}(Q^{\ast})=F^{M}_C(Q^{\ast},a)F^N_D(Q^{\ast},a)
G_{MN}(F(Q^{\ast},a)),
\]
 ($e$ is an identity element  of the group $\cal G$). 
$F^A$ are the functions by which the right action of the group $\cal G $  on a manifold $\cal P$ is realized,  
$F^{C}_{B}(Q,a)\equiv \frac{\partial F^{C}}
{\partial Q^{B}}(Q,a)$. 

$K_{\mu}$ are the Killing vector fields for the Riemannian metric 
$G_{AB}(Q)$. In (\ref{8}) they are constrained to the submanifold $\Sigma \equiv\{{\chi}^{\alpha}=0\}$, i.e.,  the components $K^A_{\mu}$ depend on $Q^{\ast}$.

In (\ref{8}), an orbit metric ${\gamma}_{\mu \nu}$ 
is defined by the  relation ${\gamma}_{\mu \nu}
=K^{A}_{\mu}G_{AB}K^{B}_{\nu}$.

The projection operators $P_{\perp}$  (depending on $Q^{\ast}$) 
is given by
\[
(P_{\perp})^{A}_{B}=\delta ^{A}_{B}-{\chi}^{\alpha}_{B}
(\chi \chi ^{\top})^{-1}{}^{\beta}_{\alpha}(\chi ^
{\top})^{A}_{\beta},
\]
 $(\chi ^{\top})^{A}_{\beta}$ is a transposed matrix to the matrix $\chi ^{\nu}_{B}\equiv \frac{\partial \chi ^{\nu}}{\partial Q^B}$, 
$(\chi ^{\top})^{A}_{\mu}=G^{AB}{\gamma}_
{\mu \nu}\chi ^{\nu}_{B}.$
This operator projects the tangent vectors onto the gauge surface $\Sigma$.

The pseudoinverse matrix ${\tilde G}^{\cal A\cal B}(Q^{\ast},a)$ to the matrix (\ref{8}), i.e., the matrix that satisfy
\begin{eqnarray*}
\displaystyle
{\tilde G}^{\cal A\cal B}{\tilde G}_{\cal B\cal C}=\left(
\begin{array}{cc}
(P_{\perp})^C_B & 0\\
0 & {\delta}^{\alpha}_{\beta}
\end{array}
\right),
\end{eqnarray*}  
 is 
\begin{equation}
\displaystyle
\left(
\begin{array}{cc}
G^{EF}N^{C}_{E}
N^{D}_{F} & G^{SD}N^C_S{\chi}^{\mu}_D
(\Phi ^{-1})^{\nu}_{\mu}{\bar v}^{\sigma}_{\nu} \\
G^{CB}{\chi}^{\gamma}_C (\Phi ^{-1})^{\beta}_{\gamma}N^D_B
{\bar v}^{\alpha}_{\beta} & G^{CB}
{\chi}^{\gamma}_C (\Phi ^{-1})^{\beta}_{\gamma}
{\chi}^{\mu}_B (\Phi ^{-1})^{\nu}_{\mu}
{\bar v}^{\alpha}_{\beta}{\bar v}^{\sigma}_{\nu}
\end{array}
\right).
\label{9}
\end{equation}
Here $(\Phi ^{-1}){}^{\beta}_{\mu}$ -- the matrix which is inverse
to
the Faddeev -- Popov matrix:
\[
(\Phi ){}^{\beta}_{\mu}(Q)=K^{A}_{\mu}(Q)
\frac{\partial {\chi}^{\beta}(Q)}{\partial Q^{A}}.
\]

 In (\ref{9}), the asymmetric projection operator $N$,
\[
N^{A}_{C}={\delta}^{A}_{C}-K^{A}_{\alpha }
(\Phi ^{-1}){}^{\alpha}_{\mu}{\chi}^{\mu}_{C},
\]
 onto the   orthogonal
to the Killing vector field subspace,
has 
the following properties:
\[
N^{A}_{B}N^{B}_{C}=N^{A}_{C},\,\,\,\,\,
(P_{\perp})^{\tilde A}_{B}N^{C}_{\tilde A}=
(P_{\perp})^{C}_{B},\,\,\,\,\,\,\,\,\,N^{\tilde A}_
{B}(P_{\perp})^{C}_{\tilde A}=N^{C}_{B}.
\]
 The matrix  ${\bar v}^{\alpha}_{\beta}(a)$ is
inverse to
the matrix ${\bar u}^{\alpha}_{\beta}(a)$. The $\det {\bar
u}^{\alpha}_{\beta}(a)$ is a density of a right invariant
measure
given on the group $\cal G$.

The determinant of the matrix (\ref{8}) is equal to
\begin{eqnarray*}
&&(\det {\tilde G}_{\cal A\cal B})=
\det G_{AB}(Q^{\ast})\det {\gamma}_{\alpha \beta}(Q^{\ast})
(\det {\chi}{\chi }^{\top})^{-1}(Q^{\ast})
(\det {\bar u}^{\mu}_{\nu}(a))^2\nonumber\\
&&\,\,\,\;\;\;\;\;\;\;\;\;\;\;\;\;\;\;\times(\det 
{\Phi}^{\alpha}_{\beta} (Q^{\ast}))^2
\det (P_{\perp})^C_B(Q^{\ast})\nonumber\\
&&\,\,\,\;\;\;\;\;\;\;\;\;\;\;\;\;\;\;=\det\Bigl((P_{\perp})^D_A \;G^{\rm H}_{DC}\,(P_{\perp})^C_B\Bigr)        \det {\gamma}_{\alpha \beta}\,\,(\det {\bar u}^{\mu}_{\nu})^2.  
\end{eqnarray*}
It does not vanish only on the surface $\Sigma$.
On this surface $\det (P_{\perp})^C_B$ is equal to  unity. 

Note also that the ``horizontal metric'' $G^{\rm H}$
is defined by the relation $G^{\rm H}_{DC}={\Pi}^{\tilde D}_D\,{\Pi}^{\tilde C}_C\,G_{{\tilde D}{\tilde C}}$ in which   ${\Pi}^{ A}_B={\delta}^A_B-K^A_{\mu}{\gamma}^{\mu \nu}K^D_{\nu}G_{DB}$ is  the projection operator. 

An original  diffusion of a scalar particle on a smooth compact Riemannian manifold $\cal P$ has been described by the backward Kolmogorov equation
 \begin{equation}
\left\{
\begin{array}{l}
\displaystyle
\left(
\frac \partial {\partial t_a}
+\frac 12\mu ^2\kappa \triangle
_{\cal P}(p_a)+\frac
1{\mu ^2\kappa m}
V(p_a)\right){\psi}_{t_b} (p_a,t_a)=0\\
{\psi}_{t_b} (p_b,t_b)=\phi _0(p_b),
\qquad\qquad\qquad\qquad\qquad (t_{b}>t_{a}).
\end{array}\right.
\label{1}
\end{equation}
In this equation $\mu ^2=\frac \hbar m$ ,
$\kappa $ is a real positive parameter,
$\triangle _{\cal P}(p_a)$  is a Laplace--Beltrami operator on manifold $\cal P$, and $V(p)$ is a group--invariant potential term. In a chart 
with the coordinate functions $Q^A=
{\varphi}^A (p)$,  the Laplace -- Beltrami operator has the standard form:
\[
\triangle _{\cal P}(Q)=G^{-1/2}(Q)\frac \partial
{\partial Q^A}G^{AB}(Q)
G^{1/2}(Q)\frac\partial {\partial Q^B},
\]
where $G=det (G_{AB})$,  $G_{AB}(Q)=G(\frac{\partial}{\partial Q^A},\frac{\partial}{\partial Q^B})$.

We have used the definition of the path integrals from \cite{Daletskii} for representing the solution of the equation (\ref{1}). This solution is written  as follows:
\begin{eqnarray}
{\psi}_{t_b} (p_a,t_a)&=&{\rm E}\Bigl[\phi _0(\eta (t_b))
\exp \{\frac 1{\mu
^2\kappa m}\int_{t_a}^{t_b}V(\eta (u))du\}\Bigr]\nonumber\\
&=&\int_{\Omega _{-}}d\mu ^\eta (\omega )
\phi _0(\eta (t_b))\exp \{\ldots \},
\label{2}
\end{eqnarray}
where  ${\eta}(t)$ is a global stochastic process on a manifold
$\cal
P$. $\Omega _{-}=\{\omega (t):\omega (t_a)=0,
\eta (t)=p_a+\omega (t)\}$  is the path space
 on this manifold. The path integral in 
 measure ${\mu}^{\eta}$ is  defined by the probability
 distribution
of a stochastic process ${\eta}(t)$.
  
Since the global semigroup determined by
the
equation (\ref{2}) 
is  defined by the limit  of the superposition of the local semigroups
\begin{equation}
\psi _{t_b}(p_a,t_a)=U(t_b,t_a)\phi _0(p_a)=
{\lim}_q {\tilde U}_{\eta}(t_a,t_1)\cdot\ldots\cdot
{\tilde U}_{\eta}(t_{n-1},t_b)
\phi _0(p_a),
\label{5}
\end{equation}
we  derive the transformation properties of
 the path integral of (\ref{2}) by studying  the local
 semigroups ${\tilde U}_{\eta}$.
These local semigroup are given by the path integrals with
 the
 integration measures determined by the local representatives
 $\eta ^A (t)$ of the global stochastic process $\eta (t)$.
The  local
processes $\eta ^{A}(t)$ are    solutions of  the stochastic
differential equations:
\begin{equation}
d\eta ^A(t)=\frac12\mu ^2\kappa G^{-1/2}\frac \partial {\partial
Q^B}(G^{1/2}G^{AB})dt+\mu \sqrt{\kappa }{\mathfrak
X}_{\bar{M}}^A(\eta
(t))
dw^{
\bar{M}}(t),
\label{3}
\end{equation}
where the matrix ${\mathfrak X}_{\bar{M}}^A$ is defined 
by the local equality  
$\sum^{n_{P}}_{\bar{
{\scriptscriptstyle K}}
\scriptscriptstyle =1}
{\mathfrak X}_{\bar{K}}^A{\mathfrak X}_
{\bar{K}}^B=G^{AB}$.\\
 (We denote the Euclidean indices by
 over--barred indices.) 

A replacement of the coordinates $Q^A$ for $(Q^{\ast}{}^A,a^{\alpha})$ does not change the path integral measures in the local semigroups as this transformation  is related to the phase space transformation of the stochastic processes.

The second step of the  reduction procedure consists of the factorization of the path integral measure. The local evolution 
given on the orbit has been separated from the evolution on the orbit space and we have come to the integral relation  between the original path integral and the corresponding path integral for the evolution on the orbit space.
The last evolution has been written in terms of the dependent coordinates. 

In particular case of the reduction performed onto the zero--momentum level, the $\lambda =0$ case of the general formula from  \cite{Storchak_2}, the integral relation between
 the path integrals for the Green's functions is 
\begin{equation}
G_{\Sigma}(Q^*_b,t_b;Q^*_a,t_a)=
\int_{\cal G}{G}_{\cal P}(p_b\theta ,t_b;
p_a,t_a)d\mu (\theta ),\;\;\;\;(Q^*=\pi_{\Sigma} (p)).
\label{intgrrelation_1}
\end{equation}
The path integral for the Green's function ${G}_{\cal P}$ may be   obtained from the path integral (\ref{2}) by choosing the delta-function as an initial function.
 The Green's function  $G_{\Sigma}$
 is presented by the following path integral 
\begin{eqnarray}
&&G_{\Sigma}(Q^*_b,t_b;Q^*_a,t_a)=\int_{ 
{{\xi}_{\Sigma}(t_a)=Q^*_a}\atop
{{\xi}_{\Sigma}(t_b)=Q^*_b}}
d\mu ^{{\xi}_{\Sigma}}\exp 
\Bigl\{\frac 1{\mu
^2\kappa m}\int_{t_a}^{t_b}
V({\xi}_{\Sigma}(u))du\Bigr\}
\nonumber\\
&&\;\;\;\;\;\;\;\;\;\;\;\;\;\;\;\;\;\times
\exp\int^{t_b}_{t_a}\Bigl\{
-\frac12{\mu}^2\kappa 
\left[(P_{\bot})^D_A\;G^{\rm H}_{DL}\,(P_{\bot})^L_B\right]
j^A_{\Roman 2}\,j^B_{\Roman 2}dt \nonumber\\
&&\;\;\;\;\;\;\;\;\;\;\;\;\;\;\;\;\;+\mu\sqrt{\kappa}\;G^{\rm H}_{DL}(P_{\bot})^D_A\,j^A_{\Roman 2}\,
  \tilde{\mathfrak X}^L_{\bar M}\,dw^{\bar M}_t
\Bigr\},
\label{pathsgm}
\end{eqnarray}
where $j_{\Roman 2}^A(Q^{*})$ 
is the projection 
of the mean curvature vector of the orbit on the submanifold $\Sigma $.
This vector has two equal representations\footnote{In \cite{Storchak_2} this vector was written with a wrong sign.}
\begin{eqnarray}
  j_{\Roman
  2}^A(Q^{*})&=&-\frac12\,G^{EU}N^A_EN^D_U
  \left[{\gamma}^{\alpha\beta}G_{CD}
  ({\tilde{\nabla}}_{K_{\alpha}}
  K_{\beta})^C\right](Q^{\ast})\nonumber\\
&=&-\frac12\,N^A_C
  \left[{\gamma}^{\alpha\beta}
  ({\tilde{\nabla}}_{K_{\alpha}}
  K_{\beta})^C\right](Q^{\ast}),
\label{j2} 
\end{eqnarray}
where
\[
({\tilde{\nabla}}_{K_{\alpha}}
  K_{\beta})^C(Q^{\ast})= 
  K^A_{\alpha}(Q^{\ast})\left.\frac{\partial}
  {\partial Q^A}K^C_{\beta}(Q)\right |_{Q=Q^{\ast}}+
  K^A_{\alpha}(Q^{\ast})K^B_{\beta}(Q^{\ast})
  {\tilde \Gamma}^C_{AB}(Q^{\ast})
    \]
  with
\[
{\tilde \Gamma}^C_{AB}(Q^{\ast})=
\frac12\
 G^{CE}(Q^{\ast})\Bigl(\frac{\partial}
 {\partial {Q^{\ast}}^A}G_{EB}(Q^{\ast})+
 \frac{\partial}
 {\partial {Q^{\ast}}^B}G_{EA}(Q^{\ast})-
 \frac{\partial}
 {\partial {Q^{\ast}}^E}G_{AB}(Q^{\ast})\Bigr).
\]

 In the path integral (\ref{pathsgm}), the path integral measure is determined by the stochastic process ${\xi} _{\Sigma}$.
Its 
local stochastic differerential equations are as follows
\begin{equation}
dQ^{*}_t{}^{\small A}={\mu}^2\kappa\biggl(-\frac12
G^{EM}N^C_EN^B_M\,{}^H{\Gamma}^A_{CB}+
j^{\small A}_{\Roman 1}\biggr)dt 
+\mu\sqrt{\kappa}
N^A_C\tilde{\mathfrak X}^C_{\bar M}dw^{\bar M}_t,
\label{sdesgm}
\end{equation}
where $j_{\Roman 1}$  is the mean curvature vector of the orbit space. The Christoffel symbols ${}^H{\Gamma}^B_{CD}$ in (\ref{sdesgm}) are defined
by the equality
\begin{equation}
G^{H}_{AB}\,\,
{}^H{\Gamma}^B_{CD}
=\frac12\left(G^{H}_{AC,D}+
G^{H}_{AD,C}-G^{H}_{CD,A}
\right),
\label{sd6}
\end{equation}
in which by the derivatives we mean the following: 
$G^{H}_{AC,D}\equiv 
\left.{{\partial G^{H}_{AC}(Q)}\over
{\partial Q^D}}\right|_{Q=Q^{*}}$.

We note that the special form of the stochastic differential equation (\ref{sdesgm})
results from the fact that 
the orbit space can be viewed  as a submanifold of the (Riemannian) manifold $({\cal P},G^H_{AB}(Q))$ with the degenerate metric $G^H_{AB}$.

The semigroup determined by the path integral (\ref{pathsgm})  acts in the space of the scalar functions given on ${\Sigma}$.
The differential generator (the Hamilton operator) of 
this semigroup is 
\begin{eqnarray}
 &&\frac12\mu ^2\kappa \left\{
 G^{CD}N^A_CN^B_D\frac{{\partial}^2}{\partial
 Q^{*}{}^A\partial
 Q^{*}{}^B}-G^{CD}N^E_CN^M_D\,{}^H{\Gamma}^A_
 {EM}\frac{\partial}{\partial
 Q^*{}^A}
\right.
 \nonumber\\
 &&+
\left.\left(j^A_{\Roman1}+j^A_{\Roman
 2}\right)\frac{\partial}{\partial Q^*{}^A}
\right\}+\frac {1}{\mu^2\kappa m}
{\tilde V}.
 \label{op2}
 \end{eqnarray} 
 
\section{Transformation of the stochastic integral}

In the integrand of the path integral (\ref{pathsgm}) there is a  term with the It\^o's stochastic  integral.
 It is not difficult to get rid of this integral by making use of 
 the It\^o's identity.
But 
first
we should rewrite the second exponent function\footnote{It is the Jacobian of the performed Girsanov transformation.} standing at  the integrand of the path integral (\ref{pathsgm}).
From the properties of  introduced projection operators it follows that  this function  can be rewritten as
\begin{eqnarray}
 {\exp}\Bigl\{-\frac18{\mu}^2{\kappa} \int\limits_{t_a}^{t_b}G^{\rm H}_{CB}{\gamma}^{\nu \sigma}({\nabla}_{K_{\nu}}K_{\sigma})^C{\gamma}^{\alpha \beta}({\nabla}_{K_{\alpha}}K_{\beta})^Bdt
& &\nonumber\\
-\,\frac12\mu \sqrt{\kappa}
\int\limits_{t_a}^{t_b}G^{\rm H}_{CD}{\gamma}^{\nu \sigma}
({\nabla}_{K_{\nu}}K_{\sigma})^C\tilde{\mathfrak X}^D_{\bar M}dw^{\bar M}_t\Bigr\}.
\label{jacgirs} 
\end{eqnarray}

By using 
the It\^o's differentiation formula 
from the  stochastic calculus, 
it can be shown 
the following equality:
\begin{eqnarray}
&&\!\!\!\!\!\!\!\!\!\!\!\!\!\!{\rm e}^{{\sigma}(Q^{\ast}(t))}=
{\rm e}^{{\sigma}(Q^{\ast}(t_a))} 
\nonumber\\
&&\times\,
{\rm e}^{\mu \sqrt{\kappa} \int\limits_{t_a}^t\frac{\partial \sigma}{\partial Q^{\ast}{}^C}B^C_{\bar M}dw^{\bar M}_t
+{\mu}^2{\kappa}\int\limits_{t_a}^t\bigl(\frac12\frac{\partial^2\sigma}{\partial Q^{\ast}{}^A Q^{\ast}{}^C}B^A_{\bar M}B^C_{\bar M}+
\frac{\partial \sigma}{\partial Q^{\ast}{}^C} a^C \bigr)dt},
\label{itoidnt_1} 
\end{eqnarray}
provided that the stochastic variable ${Q^{\ast}_t}^A$ satisfies the stochastic differential equation
\[
 d{Q^{\ast}_t}^A=a^A(Q^{\ast}_t)\,dt+B^A_{\bar M}(Q^{\ast}_t)\,dw^{\bar M}_t.
\]
Eq.(\ref{itoidnt_1}) leads to the It\^o's identity by which
\begin{eqnarray}
&&\!\!\!\!\!\!\!\!\!\!\!\!\!\!\!\!\!\!\!\!\!{\rm e}^{\mu \sqrt{\kappa} \int\limits_{t_a}^{t_b}\left(\frac{\partial \sigma}{\partial Q^{\ast}{}^C}\right)B^C_{\bar M}dw^{\bar M}_t}=\nonumber\\
&&\left(\frac{{\rm e}^{{\sigma}(Q^{\ast}(t_b))}}{{\rm e}^{{\sigma}(Q^{\ast}(t_a))}}\right)
{\rm e}^{-{\mu}^2 {\kappa} \int\limits_{t_a}^{t_b}
\left(\frac12\frac{\partial^2\sigma}{\partial Q^{\ast}{}^A\, Q^{\ast}{}^C}B^A_{\bar M}B^C_{\bar M}+
\frac{\partial \sigma}{\partial Q^{\ast}{}^C} a^C \right)dt}.
\label{itoidnt_2}
\end{eqnarray}

In order that this equality may be applied to our case, i.e., when the stochastic variable $Q^{\ast}_t$ satisfies the equation (\ref{sdesgm}),
the integrand of the stochastic integral in  (\ref{jacgirs}) must be appropriately transformed.
It may be done by making use of the identity
\[
 {\gamma}^{\sigma \mu}({\nabla}_{K_{\mu}}K_{\sigma})^E(Q^{\ast})=-\frac12G^{PE}N^A_P\Bigl({\gamma}^{\alpha \beta}\frac{\partial}{\partial Q^{\ast}{}^A} \,{\gamma}_{\alpha \beta}\Bigr)(Q^{\ast})
\]
and taking into account the following properties of the projection operators: $G^{\rm H}_{KC}G^{PC}={\Pi}^P_K$ and  ${\Pi}^P_K N^A_P=N^A_K$.
That is, in 
(\ref{itoidnt_2}) we may put 
\[
 \sigma=\frac12 {\ln}\det {\gamma}_{\alpha \beta},
\]
and $B^A_{\bar M}=N^A_L\tilde{\mathfrak X}^L_{\bar M}$. Also, we note that 
 $\frac{\partial \sigma}{\partial Q^{\ast}{}^C}=\frac12 \bigl({\gamma}^{\alpha \beta}\frac{\partial}{\partial Q^{\ast}{}^C} \,{\gamma}_{\alpha \beta}\bigr)$.

In our case,  on the right-hand side of (\ref{itoidnt_2}) we have 
the integral  with 
 the following integrand:
\begin{eqnarray}
 &&\frac14G^{\tilde A \tilde B}N^A_{\tilde A}N^B_{\tilde B}\frac{\partial}{\partial Q^{\ast}{}^A}\Bigl({\gamma}^{\alpha \beta}\frac{\partial}{\partial Q^{\ast}{}^B} \,{\gamma}_{\alpha \beta}\Bigr)\nonumber\\
&&\;\;\;\;+\frac12\Bigl(-\frac12 G^{EM}N^P_EN^S_M{}^{\rm H}{\Gamma}^A_{PS}+j^A_{\rm\Roman 1}\Bigr)\Bigl({\gamma}^{\alpha \beta}\frac{\partial}{\partial Q^{\ast}{}^A} \,{\gamma}_{\alpha \beta}\Bigr).
\label{integrand_1}
\end{eqnarray}
Since the mean curvature vector $j^A_{\rm\Roman 1}$ of the orbit space is equal to
\[
 j^A_{\rm\Roman 1}=\frac12G^{\tilde B \tilde D}N^B_{\tilde B}N^D_{\tilde D}\Bigl(N^A_{BD}+{}^{\rm H}{\Gamma}^A_{BD}-N^A_C\,{}^{\rm H}{\Gamma}^C_{BD}\Bigr),
\]
($N^A_{BD}\equiv\frac{\partial}{\partial Q^{\ast}{}^D}N^A_B$ ), 
(\ref{integrand_1}) can be transformed into
\begin{eqnarray}
 &&\frac14G^{\tilde A \tilde B}N^A_{\tilde A}N^B_{\tilde B}\frac{\partial}{\partial Q^{\ast}{}^A}\Bigl({\gamma}^{\alpha \beta}\frac{\partial}{\partial Q^{\ast}{}^B} \,{\gamma}_{\alpha \beta}\Bigr)\nonumber\\
&&\;\;\;\;+\frac14G^{\tilde B \tilde D}N^B_{\tilde B}N^D_{\tilde D}\Bigl(N^A_{BD}-N^A_C\,{}^{\rm H}{\Gamma}^C_{BD}\Bigr)\Bigl({\gamma}^{\alpha \beta}\frac{\partial}{\partial Q^{\ast}{}^A} \,{\gamma}_{\alpha \beta}\Bigr).
\label{integrand_2}
\end{eqnarray}
In the obtained expression, the second component of the sum may be further simplified.
$G^{\tilde B \tilde D}N^B_{\tilde B}N^D_{\tilde D}N^A_{BD}$ can be rewritten as 
\begin{equation}
G^{\tilde B \tilde D}N^B_{\tilde B}N^D_{\tilde D}N^A_{BD}=
G^{\tilde B \tilde D}N^D_{\tilde D}N^A_{\tilde{B} D}-G^{\tilde B \tilde D}K^B_{\mu}{\Lambda}^{\mu}_{\tilde B}N^D_{\tilde D}N^A_{BD}
\label{alf}
\end{equation}
and 
\begin{equation}
 G^{\tilde B \tilde D}N^B_{\tilde B}N^D_{\tilde D}N^A_C\,{}^{\rm H}{\Gamma}^C_{BD}=G^{\tilde B  D}N^B_{\tilde B}N^A_C\,{}^{\rm H}{\Gamma}^C_{BD}-G^{\tilde B \tilde D}N^B_{\tilde B}K^D_{\mu}{\Lambda}^{\mu}_{\tilde D}N^A_C\,{}^{\rm H}{\Gamma}^C_{BD}.
\label{bet}
\end{equation}
By using the identities $K^B_{\mu}N^A_{BD}=-K^B_{\mu D}N^A_B$ and  $N^A_C(K^C_{\alpha D}+K^B_{\alpha}\, {}^{\rm H}{\Gamma}^C_{DB})=0$,
one can show that the last terms of the equalities ({\ref{alf}) and (\ref{bet}) are equal.
Hence, they don't make  a contribution to  (\ref{integrand_2}). 
 We  obtain, therefore, the following expression for  the  integrand: 
\begin{eqnarray*}
 &&\frac14G^{\tilde A \tilde B}N^A_{\tilde A}N^B_{\tilde B}\frac{\partial}{\partial Q^{\ast}{}^A}\Bigl({\gamma}^{\alpha \beta}\frac{\partial}{\partial Q^{\ast}{}^B} \,{\gamma}_{\alpha \beta}\Bigr)\nonumber\\
&&\;\;\;\;+\frac14\Bigl(G^{\tilde B \tilde D}N^D_{\tilde D}N^A_{\tilde{B} D}-G^{\tilde B  D}N^B_{\tilde B}N^A_C\,{}^{\rm H}{\Gamma}^C_{BD}\Bigr)\Bigl({\gamma}^{\alpha \beta}\frac{\partial}{\partial Q^{\ast}{}^A} \,{\gamma}_{\alpha \beta}\Bigr).
\end{eqnarray*}
Thus, after application of  the It\^o's identity in (\ref{jacgirs}),  the  path integral reduction Jacobian can be rewriten as follows:
\begin{eqnarray}
\Bigl(\frac{{\gamma}(Q^{\ast}(t_b))}{{\gamma}(Q^{\ast}(t_a))}\Bigr)^{\frac14}
 {\exp}\Bigl\{-\frac18{\mu}^2{\kappa}\int\limits_{t_a}^{t_b}{\tilde J}dt\Bigr\},
\label{redjacob_1}
\end{eqnarray}
where
\begin{eqnarray}
&&{\tilde J}=\frac14G^{PB}N^A_BN^E_P\Bigl({\gamma}^{\mu \nu}\frac{\partial}{\partial Q^{\ast}{}^A} \,{\gamma}_{\mu \nu}\Bigr)\Bigl({\gamma}^{\alpha \beta}\frac{\partial}{\partial Q^{\ast}{}^E} \,{\gamma}_{\alpha \beta}\Bigr)\nonumber\\
&&\;\;\;\;\;\;+G^{\tilde A \tilde B}N^A_{\tilde A}N^B_{\tilde B}\frac{\partial}{\partial Q^{\ast}{}^A}\Bigl({\gamma}^{\alpha \beta}\frac{\partial}{\partial Q^{\ast}{}^B} \,{\gamma}_{\alpha \beta}\Bigr)\nonumber\\
&&\;\;\;\;\;\;+\Bigl(G^{\tilde C \tilde A}N^F_{\tilde A}N^B_{\tilde{C} F}-G^{\tilde A \tilde C}N^E_{\tilde A}N^B_M\,{}^{\rm H}{\Gamma}^M_{E \tilde C}\Bigr)\Bigl({\gamma}^{\alpha \beta}\frac{\partial}{\partial Q^{\ast}{}^B} \,{\gamma}_{\alpha \beta}\Bigr).
\label{jacobintgrand}
\end{eqnarray}
The main task of the present paper is to get the geometrical representation for the integrand ${\tilde J}$. It will be done with the formula for the Riemannian curvature scalar of $\cal P$.

\section{The scalar curvature of the bundle}

The scalar curvature of 
the Riemannian manifold $\cal P$
will be calculated by using  the special nonholonomic basis. This basis
 generalizes the horizontal lift basis considered 
in \cite{Cho}. It consists of the horizontal vector fields $H_A$ and the left-invariant vector fields $L_{\alpha}=v^{\mu}_{\alpha}(a)\frac{\partial}{\partial a^{\mu}}$.
The vector fields $L_{\alpha}$ have the standard commutation relations 
\[
[L_{\alpha},L_{\beta}]=c^{\gamma}_{\alpha \beta} L_{\gamma},
\]
where the $c^{\gamma}_{\alpha \beta}$ are the structure constant of the group $\cal G$.

The vector fields $H_A$ are given by
\[
 H_A=N^E_A(Q^{\ast}) \left(\frac{\partial}{\partial Q^{\ast}{}^E}-{\tilde {\mathscr A}}^{\alpha}_E\,L_{\alpha}\right),
\]
where ${\tilde{\mathscr A} }^{\alpha}_E(Q^{\ast},a)={\bar{\rho}}^{\alpha}_{\mu}(a)\,{\mathscr A}^{\mu}_E(Q^{\ast})$. 
The matrix ${\bar{\rho}}^{\alpha}_{\mu}$ is inverse to the matrix ${\rho}_{\alpha}^{\beta}$ of the adjoint representation of the group $\cal G$, and  ${\mathscr A}^{\nu}_P={\gamma}^{\nu\mu}K^R_{\mu}\,G_{RP}$ is the ``mechanical'' connection  defined on our principal fibre bundle.

The commutation relation  of the horizontal vector fields are 
\[
 [H_C,H_D]=({\Lambda}^{\gamma}_CN^P_D-{\Lambda}^{\gamma}_DN^P_C)K^{S}_{{\gamma} P}\,H_S-N^E_CN^P_D\,\tilde{\mathcal F}^{\alpha}_{EP}L_{\alpha},
\]
where ${\Lambda}^{\gamma}_D=({\Phi}^{-1})^{\gamma}_{\mu}\,{\chi}^{\mu}_D$.
The curvature $\tilde{\mathcal F}^{\alpha}_{EP}$ of the connection ${\tilde{\mathscr A}}$ is given by
\[
\tilde{\mathcal F}^{\alpha}_{EP}=\displaystyle\frac{\partial}{\partial Q^{\ast}{}^E}\,\tilde{\mathscr A}^{\alpha}_P- 
\frac{\partial}{\partial {Q^{\ast}}^P}\,\tilde{\mathscr A}^{\alpha}_E
+c^{\alpha}_{\nu\sigma}\, \tilde{\mathscr A}^{\nu}_E\,
\tilde{\mathscr A}^{\sigma}_P,
\]
($\tilde{\mathcal F}^{\alpha}_{EP}({Q^{\ast}},a)={\bar{\rho}}^{\alpha}_{\mu}(a)\,{\mathcal F}^{\mu}_{EP}(Q^{\ast})\,$). In calculation of the commutation relation we have used the following equality:
\[
L_{\alpha}\, {\tilde {\mathscr A}}^{\lambda}_E=-c^{\lambda}_{\alpha \mu}
\,{\tilde {\mathscr A}}^{\mu}_E,
\]
which comes  from the equation satisfied by ${\rho}$:  $L_{\alpha}\,{\rho}^{\gamma}_{\beta}=c^{\mu}_{\alpha \beta}\,{\rho}^{\gamma}_{\mu}$.

The above commutation relation may be also written as follows:
\[
 [H_C,H_D]={\mathscr C}^A_{CD}\,H_A+{\mathscr C}^{\alpha}_{CD}L_{\alpha},
\]
where the structure constants of the nonholonomic basis are
\[{\mathscr C}^A_{CD}=({\Lambda}^{\gamma}_CK^A_{\gamma D}-{\Lambda}^{\gamma}_DK^{A}_{{\gamma} C})\]
and
\[{\mathscr C}^{\alpha}_{CD}=-N^S_CN^P_D\,\tilde{\mathcal F}^{\alpha}_{SP}\,.
\]

In our  basis, $L_{\alpha}$
commutes with $H_A$\,:
\[
[H_A,L_{\alpha}]=0.
\]

In the horizontal lift basis, the metric  (\ref{8}) can be written as 
\begin{equation}
\displaystyle
{\check G}_{\cal A\cal B}=
\left(
\begin{array}{cc}
G^{\rm H}_{AB} & 0 \\
0 & \tilde{\gamma }_{\alpha \beta }
\end{array}
\right),
\label{metric}
\end{equation}
with
\[
 {\tilde G}(H_A,H_B)\equiv G^{\rm H}_{AB}(Q^{\ast}), \;\;\;\; 
{\tilde{G}}(L_{\alpha},L_{\beta})\equiv\tilde{\gamma }_{\alpha \beta }(Q^{\ast},a)={\gamma}_{{\alpha}'{\beta}'}(Q^{\ast})\, {\rho}^{{\alpha}'}_{\alpha}(a)\,
{\rho}^{{\beta}'}_{\beta}(a).
\]

The dual basis  is given by the following one-forms:
\begin{eqnarray*}
&&{\omega}^A=({\rm P}_{\bot})^A_S\, dQ^{\ast}{}^S,\nonumber\\
&&{\omega}^{\alpha}=u^{\alpha}_{\mu}da^{\mu}+{\tilde{\mathscr A}}^{\alpha}_E \,({\rm P}_{\bot})^E_S\, dQ^{\ast}{}^S,
 \end{eqnarray*}
for which  ${\omega}^{A}(H_B)=N^A_B$,   ${\omega}^{\alpha}(H_A)=0$, and ${\omega}^{\alpha}(L_{\beta})={\delta}^{\alpha}_{\beta}$.

The pseudoiverse matrix ${\check G}^{{\mathcal A}{\mathcal B}}$ to the matrix (\ref{metric}) is defined by
\begin{eqnarray*}
\displaystyle
{\check G}^{\cal A\cal B}=
\left(
\begin{array}{cc}
G^{EF}N^A_EN^B_F & 0 \\
0 & \tilde{\gamma }^{\alpha \beta }
\end{array}
\right),
\end{eqnarray*}
with
\[
{\tilde{G}}({\omega}^A,{\omega}^B)\equiv G^{EF}N^A_EN^B_F,\;\;
{\tilde{G}}({\omega}^{\alpha},{\omega}^{\beta})\equiv 
\tilde{\gamma}^{{\alpha}{\beta}}={\gamma}^{{\alpha}'{\beta}'} {\bar\rho}^{{\alpha}}_{{\alpha}'}
{\bar\rho}^{{\beta}}_{{\beta}'},\;\;{\tilde{G}}({\omega}^A,{\omega}^{\alpha})=0.
 \]

The orthogonality condition  is
\begin{eqnarray*}
\displaystyle
{\check G}^{\mathcal A\mathcal B}{\check G}_{\mathcal B\mathcal C}=\left(
\begin{array}{cc}
N^A_C & 0\\
0 & {\delta}^{\alpha}_{\beta}
\end{array}
\right).
\end{eqnarray*}  

\subsection{The Christoffel symbols}
The computation of the connection coefficients 
$\check{\Gamma}^{{\cal D}}_{\mathcal A \mathcal B}$ in the nonholonomic basis 
 will be preformed by using the following formula: 
\begin{eqnarray}
 2\,\check{\Gamma}^{{\cal D}}_{\mathcal A \mathcal B}\,{\tilde G}(
{\partial}_{\mathcal D},{\partial}_{\mathcal C})={\partial}_{\mathcal A}\,{\tilde G}({\partial}_{\mathcal B},{\partial}_{\mathcal C})+{\partial}_{\mathcal B}\,{\tilde G}({\partial}_{\mathcal A},{\partial}_{\mathcal C})-{\partial}_{\mathcal C}\,{\tilde G}({\partial}_{\mathcal A},{\partial}_{\mathcal B})& &\nonumber\\
 -{\tilde G}({\partial}_{\mathcal A},[{\partial}_{\mathcal B},{\partial}_{\mathcal C}])-{\tilde G}({\partial}_{\mathcal B},[{\partial}_{\mathcal A},{\partial}_{\mathcal C}])+{\tilde G}({\partial}_{\mathcal C},[{\partial}_{\mathcal A},{\partial}_{\mathcal B}]),
\label{christ_0}
\end{eqnarray}
where the terms of the form  ${\partial}_{\mathcal A}\tilde G$ denote the corresponding  directional derivatives.  

First we consider the calculation of the coordinate components
${\check {\Gamma}}^A_{BC}$. In  this  case (\ref{christ_0}) can be  rewritten as 
\[
2 {\check {\Gamma}}^D_{AB}\,G^{\rm H}_{DC}=H_AG^{\rm H}_{BC}+H_BG^{\rm H}_{AC}-H_CG^{\rm H}_{AB}-G^{\rm H}_{AP}{\mathscr C}^P_{BC}-G^{\rm H}_{BP}{\mathscr C}^P_{AC}+G^{\rm H}_{CP}{\mathscr C}^P_{AB}.
\]
Replacing structure constant ${\mathscr C}^A_{BC}$ by their explicit expressions and performing the necessary transformations, we get  
\begin{equation}
 {\check {\Gamma}}^D_{AB}\,G^{\rm H}_{DC}=N^E_A\, {}^{\rm H}{\Gamma}_{BEC}-N^E_B \,{}^{\rm H}{\Gamma}_{ACE}+N^E_C \,{}^{\rm H}{\Gamma}_{ABE}+{}^{\rm H}{\Gamma}_{ACB}-{}^{\rm H}{\Gamma}_{BAC},
\label{christ_1}
\end{equation}
where
\[
 {}^{\rm H}{\Gamma}_{ABC}=G^{\rm H}_{AC,B}+G^{\rm H}_{BC,A}-G^{\rm H}_{AB,C}.
\]
Eq.(\ref{christ_1}) was obtained by  making  use of the equality  
$G^{\rm H}_{AP}K^P_{\mu C}=-K^P_{\mu}G^{\rm H}_{AP,C}$, which can be  derived   by means of the differentiation (with respect to $Q^{\ast}$) of the relation $G^{\rm H}_{AP}K^P_{\mu}=0$.

From the Killing identity for the metric $G_{AB}$
it follows that
\[
K^E_{\mu}G^{\rm H}_{AB,E}=0.
\]
Taking this equality into account   we transform eq.(\ref{christ_1}) into 
\[
 G^{\rm H}_{DC}\,{\check {\Gamma}}^D_{AB}=N^E_A\,{}^{\rm H}{\Gamma}_{BEC}. 
\]
Multiplying the both sides of this equation on $G^{SF}N^P_SN^C_F$
and 
using the identity $G^{MS}N^A_M\,{}^{\rm H}{\Gamma}_{CDS}=N^A_S\,{}^{\rm H}{\Gamma}^S_{CD}$, we get
\[
N^P_S\,{\check {\Gamma}}^S_{AB}=N^P_KN^E_A\,{}^{\rm H}{\Gamma}^K_{BE}. 
\]

Hence, we can write 
\[
{\check {\Gamma}}^D_{AB}=N^E_A\,{}^{\rm H}{\Gamma}^D_{BE}
\]
(modulo the terms $X^S_{ABC}$ for which  $N^P_SX^S_{ABC}=0$).

The calculation of other non-vanishing  
coordinate components of the Christoffel connection 
$\check{\Gamma}^{{\cal D}}_{\mathcal A \mathcal B}$ on $\cal P$
leads to the following result: 
\begin{eqnarray*}
&&{\check {\Gamma}}^{\mu}_{AB}=-\frac12N^E_AN^F_B\,\tilde{\mathcal F}^{\mu}_{EF},\nonumber\\
&&{\check {\Gamma}}^{P}_{\alpha B}=\frac12 G^{PS}N^F_SN^E_B\,\tilde
{\mathcal F}^{\mu}_{EF}{\tilde \gamma}_{\mu\alpha},\nonumber\\
&&{\check {\Gamma}}^{P}_{A\beta}=\frac12 G^{PS}N^F_SN^E_A\,\tilde
{\mathcal F}^{\mu}_{EF}{\tilde \gamma}_{\mu\beta},\nonumber\\
&&{\check {\Gamma}}^{P}_{\alpha\beta}=-\frac12\,G^{PS}H_S{\tilde \gamma}_{\alpha \beta}=
-\frac12 G^{PS}N^E_S\,{\tilde \mathscr D}_E{\tilde \gamma}_{\alpha\beta},\nonumber\\
&&{\check{\Gamma}}^{\mu}_{\alpha B}   
=\frac12{\tilde \gamma}^{\mu\nu}H_B{\tilde \gamma}_{\alpha\nu}=
\frac12{\tilde\gamma}^{\mu\nu}N^E_B
\,{\tilde \mathscr D}_E{\tilde \gamma}_{\alpha\nu},\nonumber\\
&&{\check{\Gamma}}^{\mu}_{A\beta}   
=\frac12{\tilde \gamma}_{\mu\nu}H_A{\tilde \gamma}_{\beta\nu}=
\frac12{\tilde\gamma}^{\mu\nu}N^E_A
\,{\tilde \mathscr D}_E{\tilde \gamma}_{\beta \nu},
\nonumber\\
&&{\check{\Gamma}}^{\mu}_{\alpha\beta}=\frac12{\tilde \gamma}^{\mu\nu}(c^{\sigma}_{\alpha\beta}{\tilde \gamma}_{\sigma\nu}-c^{\sigma}_{\nu\beta}{\tilde \gamma}_{\alpha\sigma}-c^{\sigma}_{\nu\alpha}{\tilde \gamma}_{\beta\sigma}).
\end{eqnarray*}
In these formulae
the covariant derivatives 
are given as follows:
\[
{\tilde \mathscr D}_E{\tilde \gamma}_{\alpha\beta}=
\Bigl(\frac{\partial}{\partial Q^{\ast}{}^E}{\tilde \gamma}_{\alpha\beta}-c^{\sigma}_{\mu\alpha}{\tilde\mathscr A}^{\mu}_E{\tilde \gamma}_{\sigma\beta}-c^{\sigma}_{\mu\beta}{\tilde\mathscr A}^{\mu}_E{\tilde \gamma}_{\sigma\alpha}\,\Bigr).
\]
\subsection{The Ricci curvature tensor}
To  evaluate the components $ \check R_{\mathcal A \mathcal C}$ of the  the Ricci curvature tensor of the metric (\ref{metric}) relative to our nonholonomic basis we
make use of the following formula: 
\begin{equation}
\check R_{\mathcal A \mathcal C}={\partial}_{\mathcal A}\check
{\Gamma}^{\mathcal P}_{\mathcal P \mathcal C}- {\partial}_{\mathcal P}\check{\Gamma}^{\mathcal P}_{\mathcal A \mathcal C}+\check{\Gamma}^{\mathcal D}_{\mathcal P \mathcal C}\check{\Gamma}^{\mathcal P}_{\mathcal A \mathcal D}-\check{\Gamma}^{\mathcal E}_{\mathcal A \mathcal C}\check{\Gamma}^{\mathcal P}_{\mathcal P \mathcal E}-
{\mathscr C}^{\mathcal E}_{\mathcal A \mathcal P}\check{\Gamma}^{\mathcal P}_{\mathcal E \mathcal C}.
\label{Ricci_0}
\end{equation}
For the components  $\check R_{A C}$ this formula
can be written as 
\begin{eqnarray}
 {\check R}_{AC}&=&H_A {\check {\Gamma}}^M_{MC}
-H_M {\check {\Gamma}}^M_{AC}+{\check {\Gamma}}^K_{MC}{\check {\Gamma}}^M_{AK}-{\check{\Gamma}}^E_{AC}{\check{\Gamma}}^K_{KE}
-{\mathscr C}^K_{AM}{\check{\Gamma}}^M_{KC}\nonumber\\
&+&H_A {\check {\Gamma}}^{\mu}_{\mu C}-L_{\alpha} {\check {\Gamma}}^{\alpha}_{AC}+{\check {\Gamma}}^{\mu}_{MC}{\check {\Gamma}}^{M}_{A\mu}+{\check {\Gamma}}^K_{\mu C}{\check {\Gamma}}^{\mu}_{AK}+{\check {\Gamma}}^{\nu}_{\mu C} {\check {\Gamma}}^{\mu}_{A\nu}
-{\check{\Gamma}}^E_{A C}{\check{\Gamma}}^{\nu}_{\nu E}\nonumber\\
&-&{\check{\Gamma}}^{\mu}_{AC}{\check{\Gamma}}^K_{K\mu}-{\check{\Gamma}}^{\mu}_{AC}{\check{\Gamma}}^{\nu}_{\nu\mu}-{\mathscr C}^{\alpha}_{AM}{\check{\Gamma}}^M_{\alpha C}.
\label{Ricci_1}
\end{eqnarray}

First of all, let us consider 
the expression standing at the first line of the right--hand side of
(\ref{Ricci_1}).
By using obtained Christoffel symbols  ${\check {\Gamma}}$ and the coefficients ${\mathscr C}^{K}_{AM}$,  we present this expression  as follows: 
\[
 N^S_AN^E_M\left(\frac{\partial}{\partial Q^{\ast}{}^S}{}^{\rm H}{\Gamma}^M_{CE}-\frac{\partial}{\partial Q^{\ast}{}^E}{}^{\rm H}{\Gamma}^M_{CS}+{}^{\rm H}{\Gamma}^K_{CE}\,{}^{\rm H}{\Gamma}^M_{KS}-{}^{\rm H}{\Gamma}^P_{CS}\,{}^{\rm H}{\Gamma}^M_{PE}\right).
\]
It
may also be rewritten as
\[
  N^S_AN^E_M\,{}^{\rm H}R^{\;\;\;\;\;\;\;M}_{SEC},
\]
where by ${}^{\rm H}R^{\;\;\;\;\;\;\;M}_{SEC}$ we denote the 
 expression  which looks like the Riemann  curvature tensor of the manifold with the degenerate metric $G^{\rm H}_{AB}$.

We note that our Christoffel symbols ${}^{\rm H}{\Gamma}^M_{CS}$ are defined up to the terms  $T^M_{CS}$ that satisfy  the equality $G^{\rm H}_{AB}T^B_{CD}=0$. Therefore, we are allowed to neglect the terms that are directly proportional to $K^B_{\alpha}$.
Since in our case
\[
 N^E_M\,{}^{\rm H}R^{\;\;\;\;\;\;\;M}_{SEC}={}^{\rm H}R^{\;\;\;\;\;\;\;M}_{SMC}-K^E_{\alpha}{\Lambda}^{\alpha}_M\,{}^{\rm H}R^{\;\;\;\;\;\;\;M}_{SEC},
\]
the contribution to  the Ricci curvature $\check R_{AC}$ coming  from the second 
 term of the previous expression may be omitted. 
The first term will give us 
$N^S_A\,{}^{\rm H}R^{\;\;\;\;\;\;\;M}_{SMC}$.

The remaining terms of ${\check R}_{AC}$ are presented by  the following  expressions:
\begin{eqnarray*}
H_A\,{}^{\rm H}{\check{\Gamma}}^{\nu}_{\nu C}&=&
\frac12H_A({\tilde \gamma}^{\mu\nu}H_C{\tilde \gamma}_{\mu\nu})\equiv
\frac12N^F_A\frac{\partial}{\partial Q^{\ast}{}^F}\left({\gamma}^{\mu\nu}N^E_C\frac{\partial}{\partial Q^{\ast}{}^E}{\gamma}_{\mu\nu}\right);\nonumber\\
L_{\alpha}{\check{\Gamma}}^{\alpha}_{AC}&=&-\frac12N^E_AN^P_C\,L_{\alpha}\tilde{\mathcal F}^{\alpha}_{EP};\nonumber\\
 {\check{\Gamma}}^{\nu}_{MC}{\check{\Gamma}}^{M}_{A\nu }&=&-\frac14(G^{MS}N^E_MN^P_S)\,N^F_CN^Q_A\,\tilde{\mathcal F}^{\mu}_{EF}\tilde{\mathcal F}^{\nu}_{QP}{\tilde \gamma}_{\mu\nu};\nonumber\\
{\check{\Gamma}}^{K}_{\nu C}{\check{\Gamma}}^{\nu}_{AK}&=&-\frac14(G^{KS}N^F_SN^R_K)\,N^E_AN^P_C\,\tilde{\mathcal F}^{\nu}_{PF}\tilde{\mathcal F}^{\mu}_{ER}{\tilde \gamma}_{\mu\nu};\nonumber\\
{\check{\Gamma}}^{\mu}_{\alpha C}{\check{\Gamma}}^{\alpha}_{A\mu}&=&
\frac14({\tilde \gamma}^{\mu\nu}H_C{\tilde \gamma}_{\alpha\nu})\,
({\tilde \gamma}^{\alpha\beta}H_A{\tilde \gamma}_{\mu\beta})\nonumber\\
&=&\frac14
{\tilde \gamma}^{\mu \nu}N^E_C\bigl({\tilde \mathscr D}_E{\tilde\gamma}_{\alpha\nu}\bigr)\,{\tilde \gamma}^{\alpha \beta}N^F_A\bigl({\tilde \mathscr D}_F{\tilde\gamma}_{\mu\beta}\bigr);
\nonumber\\
 {\check{\Gamma}}^{E}_{AC}{\check{\Gamma}}^{\nu}_{\nu E}&=&\frac12N^P_A\,{}^{\rm H}{{\Gamma}}^{E}_{CP}\,\bigl({\tilde \gamma}^{\mu\nu}H_E{\tilde \gamma}_{\mu\nu}\bigr).
\end{eqnarray*}
In derivation of these terms we have used the identity $c^{\sigma}_{\sigma\mu}=0$, which is valid for the compact semisimple Lie group.

Collecting the parts of ${\check R}_{AC}$, we get
\begin{eqnarray}
{\check R}_{AC}&=&N^S_AN^E_M\;{}^{\rm H}R_{SEC}^{\;\;\;\;\;\;\;\;M}-\frac12N^E_AN^F_C\,L_{\alpha}\tilde{\mathcal F}^{\alpha}_{EF}+\frac12 N^E_AN^F_C\,\tilde{\mathcal F}^{\alpha}_{EF}{\check  \Gamma}^{\nu}_{\nu \alpha}
\nonumber\\
&+&\frac12 (G^{MS}N^P_MN^F_S)\,N^E_AN^R_C\,\tilde{\mathcal F}^{\alpha}_{EP}\tilde{\mathcal F}^{\mu}_{RF}{\tilde  \gamma}_{\mu \alpha}-\frac12N^P_A\,{}^{\rm H}{\Gamma}^E_{CP}({\tilde \gamma}^{\mu \nu}H_E\,{\tilde \gamma}_{\mu \nu})
\nonumber\\
&+&\frac12 H_A\,\bigl({\tilde \gamma}^{\mu \nu}H_C\,{\tilde \gamma}_{\mu \nu}\bigr)+\frac14\bigl({\tilde \gamma}^{\mu \nu}H_C\,{\tilde \gamma}_{\alpha \nu}\bigr)\bigl({\tilde \gamma}^{\alpha \beta}H_A\,{\tilde \gamma}_{\mu \beta}\bigr).
\label{Ricci_2}
\end{eqnarray}

The components $\check R_{\alpha \beta}$ of the Ricci curvature 
$\check R_{\mathcal A \mathcal C}$ are defined as 
\[
 \check R_{\alpha \beta}=\hat{\partial}_{\alpha}{\check \Gamma}^{\mathcal K}_{\mathcal K \beta}-\hat{\partial}_{\mathcal K}{\check \Gamma}^{\mathcal K}_{\alpha \beta}+{\check \Gamma}^{\mathcal E}_{\mathcal K \beta}{\check \Gamma}^{\mathcal K}_{\alpha \mathcal E}-{\check \Gamma}^{\mathcal E}_{\alpha \beta}{\check \Gamma}^{\mathcal K}_{\mathcal K \mathcal E}-{\mathscr C}^{\mathcal E}_{\alpha \mathcal K}{\check \Gamma}^{\mathcal K}_{\mathcal E \beta}.
\]
In our case $\check R_{\alpha \beta}$ are given by the following formula:
\begin{eqnarray}
 &&\check R_{\alpha \beta}= \tilde R_{\alpha \beta}+\frac14\bigl(G^{ES}N^F_SN^B_E\bigr)\,\bigl(G^{MQ}N^P_MN^A_Q\bigr)\,{\tilde \gamma}_{\mu \beta}{\tilde \gamma}_{\nu \alpha}\,{\tilde\mathcal F}^{\mu}_{PF}{\tilde\mathcal F}^{\nu}_{BA}\nonumber\\
&&\;\;\;\;\;+\frac12\,H_M\bigl(G^{MS}H_S{\tilde \gamma}_{\alpha \beta}\bigr)
-\frac14\bigl({\tilde \gamma}^{\sigma \nu}H_M{\tilde \gamma}_{\sigma \beta}\bigr)\,\bigl(G^{MS}H_S{\tilde \gamma}_{\alpha \nu}\bigr)\nonumber\\
&&\;\;\;\;\;-\frac14\,\bigl(G^{ES}H_S{\tilde \gamma}_{\nu \beta}\bigr)\,\bigl({\tilde \gamma}^{\sigma \nu}H_E{\tilde \gamma}_{\sigma \alpha}\bigr)
+\frac12\,N^Q_M\,{}^{\rm H}{ \Gamma}^M_{EQ}\bigl(G^{ES}H_S{\tilde \gamma}_{\alpha \beta}\bigr)
\nonumber\\
&&\;\;\;\;\;+\frac14\bigl(G^{ES}H_S{\tilde \gamma}_{\alpha \beta}\bigr)\bigl({\tilde \gamma}^{\mu \nu}H_E{\tilde \gamma}_{\mu \nu}\bigr),
\label{Ricci_3}
\end{eqnarray}
in which by $\tilde R_{\alpha \beta}$ we denote the Ricci curvature of the manifold with the Riemannian metric ${\tilde \gamma}_{\mu \nu}$.
 
\subsection{The calculation of the scalar curvature}
In the horizontal lift  basis the scalar curvature of the original manifold $\mathcal P$ is defined by 
\begin{equation}
  R_{\mathcal P}=G^{\tilde A\tilde C}N^A_{\tilde A}N^C_{\tilde C}\,
{\check R}_{AC}+{\tilde\gamma}^{\alpha \beta}\check R_{\alpha \beta}.
\label{scalcurv}
\end{equation}
Notice that by the symmetry argument the second and the third terms 
of (\ref{Ricci_2}) will not make the contributions into $ R_{\mathcal P}$.

First we consider the contribution to $ R_{\mathcal P}$ 
which is obtained
from the  terms belonging to ${\check R}_{AC}$ and ${\check R}_{\alpha \beta}$
that are given with a single multiplier 
$({\tilde \gamma}^{\mu \nu}{\tilde \mathscr D}_A{\tilde \gamma}_{\mu \nu})$. 
In (\ref{Ricci_2}), they are the fifth and  sixth terms. 
 
It can be shown that the sixth term of ${\check R}_{AC}$ leads to  the following contribution:
\begin{equation}
\frac12G^{\tilde A\tilde C}N^E_{\tilde A}N^B_{\tilde C}\,{\tilde \mathscr D}_E\bigl({\tilde \gamma}^{\mu \nu}{\tilde \mathscr D}_B{\tilde \gamma}_{\mu \nu}\bigr)+\frac12G^{\tilde A\tilde C}N^E_{\tilde A}N^C_{\tilde C}N^A_{CE}\bigl({\tilde \gamma}^{\mu \nu}{\tilde \mathscr D}_A{\tilde \gamma}_{\mu \nu}\bigr).
\label{contrib_6}
\end{equation}
Combining (\ref{contrib_6}) with the contribution of the fifth term to $ R_{\mathcal P}$ , we get
\begin{equation}
 \frac12 G^{\tilde A\tilde C}N^{P'}_{\tilde A}N^{C'}_{\tilde C}
\bigl(N^A_{C'P'}-N^A_E\,{}^{\rm H}{\Gamma}^A_{C'P'}\bigr)\,({\tilde \gamma}^{\mu \nu}{\tilde \mathscr D}_A{\tilde \gamma}_{\mu \nu}).
\label{terms_1rac}
\end{equation}

Now we will calculate the corresponding contribution to $ R_{\mathcal P}$ originated from the terms of
$\check R_{\alpha \beta}$. But before proceeding to  calculation these terms
 must be  transformed.
First we   rewrite the third term of (\ref{Ricci_3})  as
\begin{equation}
H_M\bigl(G^{MS}\bigr)\bigl(H_S{\tilde \gamma}_{\alpha \beta}\bigr)+G^{MS}N^E_MN^P_{SE}{\tilde \mathscr D}_P{\tilde \gamma}_{\alpha \beta} +G^{MS}N^E_MN^P_S
{\tilde \mathscr D}_E\bigl({\tilde \mathscr D}_P{\tilde \gamma}_{\alpha \beta}\bigr).
\label{term_3}
\end{equation}
 Then, differentiating the identity 
\[
 N^A_L=G^{\rm H}_{LF}\bigl(G^{MS}N^F_MN^A_S\bigr)
\]
with respect to $Q^{\ast}{}^E$, we get
 the following  equality:
\begin{eqnarray*}
 &&\bigl(G^{MS}\bigr)_{,\,E}N^E_MN^A_S=\nonumber\\
&&\;\;\;\;\;-\,\bigl(G^{MS}N^F_MN^A_S\bigr)N^E_B\,{}^{\rm H}{\Gamma}^B_{FE}-G^{LU}N^A_BN^E_U\,{}^{\rm H}{\Gamma}^B_{LE}-G^{MS}N^F_{ME}N^E_FN^A_S.
\end{eqnarray*}
By making use of this equality, it can be shown that the contribution of (\ref{term_3}) into $R_{\mathcal P}$
are given by the following expression:
\begin{eqnarray}
&&\frac12\bigl(- G^{MS}N^F_MN^A_SN^E_B\,{}^{\rm H}{\Gamma}^B_{FE}-G^{LU}N^A_BN^E_U\,{}^{\rm H}{\Gamma}^B_{LE}-G^{MS}N^F_{ME}N^E_FN^A_S\nonumber\\
&&+G^{MS}N^E_{M}N^A_{SE}+G^{FS}N^A_SN^E_B\,{}^{\rm H}{\Gamma}^B_{EF}\bigr)({\tilde \gamma}^{\mu \nu}{\tilde \mathscr D}_A{\tilde \gamma}_{\mu \nu}).
\label{term_31}
\end{eqnarray}
Using  $N^F_M={\delta}^F_M-K^F_{\mu}{\Lambda}^{\mu}_M$ for the projector $N^F_M$  in the  first term of (\ref{term_31}),
we see that the part of the first term and the last term of this expression are mutually cancelled. 
Besides,   grouping  the third term in (\ref{term_31})
with the remnant of the first term, we also come  
to zero. It take place because of the identity 
\[
N^C_P(K^P_{\alpha E}+K^F_{\alpha}\,{}^{\rm H}{\Gamma}^P_{FE})=0 
\]
which is derived from the Killing relation for the horizontal metric $G^{\rm H}_{AB}$:
\[
 K^E_{\mu B}G^{\rm H}_{AE}+K^E_{\mu A}G^{\rm H}_{BE}+K^E_{\mu }G^{\rm H}_{AB,E}=0.
\]
 Notice that before using the identity in (\ref{term_31}),  one should make a replacement of $N^E_FN^F_{ME}$ for $-N^E_PK^P_{\nu E}{\Lambda}^{\nu}_M$ in the third term. 

The second and  fourth terms of (\ref{term_31})
give us 
\begin{equation}
 \frac12\,G^{LU}N^E_U\bigl(N^A_{LE}-N^A_B\,{}^{\rm H}{\Gamma}^B_{LE}\bigr)({\tilde \gamma}^{\mu \nu}{\tilde \mathscr D}_A{\tilde \gamma}_{\mu \nu}).
\label{terms_1ralbet}
\end{equation}
The obtained expression is the contribution to $ R_{\mathcal P}$ given by  the terms of  $\check R_{\alpha \beta}$.
  
Taking a sum of (\ref{terms_1rac}) and (\ref{terms_1ralbet}), we get the contribution to $ R_{\mathcal P}$ which is obained  from 
$\check R_{AC}$ and $\check R_{\alpha \beta}$:
\begin{eqnarray}
 G^{{\tilde A}C'}N^{P'}_{\tilde A}\bigl(N^A_{C'P'}-N^A_E
{}^{\rm H}{\Gamma}^B_{C'P'}\bigr)
({\tilde \gamma}^{\mu \nu}{\tilde \mathscr D}_A{\tilde \gamma}_{\mu \nu})& &\nonumber\\
 \equiv G^{\tilde A C'}N^{P'}_{\tilde A}\bigl({}^{\rm H}{\nabla}_{P'}N^A_{C'}\bigr)({\tilde \gamma}^{\mu \nu}{\tilde \mathscr D}_A{\tilde \gamma}_{\mu \nu}).
\label{term_1r_P}
\end{eqnarray}

Finally, we  consider  the contribution to $ R_{\mathcal P}$ which is obtained  from the terms belonging to
 ${\check R}_{AC}$ and ${\check R}_{\alpha \beta}$, and  
containing
the product of  two multipliers of the 
aforementioned kind. 

The terms of  ${\check R}_{AC}$ give   the following expression as the contribution to $ R_{\mathcal P}$:
\[
\frac12G^{\tilde A\tilde C}N^E_{\tilde A}N^B_{\tilde C}\,{\tilde \mathscr D}_E\bigl({\tilde \gamma}^{\mu \nu}{\tilde \mathscr D}_B{\tilde \gamma}_{\mu \nu}\bigr)
+\frac14G^{\tilde A\tilde C}N^A_{\tilde A}N^C_{\tilde C}\,\bigl({\tilde \gamma}^{\mu \nu}{\tilde \mathscr D}_C{\tilde \gamma}_{\alpha \nu}\bigr)\bigl({\tilde \gamma}^{\alpha \beta}{\tilde \mathscr D}_A{\tilde \gamma}_{\mu \beta}\bigr).
\] 
The contribution from the terms of ${\check R}_{\alpha \beta}$             can be presented as follows:
\begin{eqnarray*}
 &&\frac12G^{MS}N^E_MN^P_S{\tilde  \gamma}^{\alpha \beta}
{\tilde \mathscr D}_E\bigl({\tilde \mathscr D}_P{\tilde \gamma}_{\alpha \beta}\bigr)-\frac14G^{MS}N^E_MN^P_S
\bigl({\tilde \gamma}^{\nu \sigma}{\tilde \mathscr D}_E{\tilde \gamma}_{\sigma \beta}\bigr)\bigl({\tilde \gamma}^{\alpha \beta}{\tilde \mathscr D}_P{\tilde \gamma}_{\alpha \nu}\bigr)
\nonumber\\
&&-\frac14G^{ES}N^P_SN^R_E\biggl[
\bigl({\tilde \gamma}^{\alpha \beta}{\tilde \mathscr D}_P{\tilde \gamma}_{\nu \beta}\bigr)\bigl({\tilde \gamma}^{\nu \sigma}{\tilde \mathscr D}_R{\tilde \gamma}_{\alpha \sigma}\bigr)
-\frac14\bigl({\tilde \gamma}^{\alpha \beta}{\tilde \mathscr D}_P{\tilde \gamma}_{\alpha \beta}\bigr)\bigl({\tilde \gamma}^{\mu \nu}{\tilde \mathscr D}_R{\tilde \gamma}_{\mu \nu}\bigr)\biggr].
\end{eqnarray*}
Replacing the first term of the last expression with the help of the equality
\[
 {\tilde  \gamma}^{\alpha \beta}
{\tilde \mathscr D}_E\bigl({\tilde \mathscr D}_P{\tilde \gamma}_{\alpha \beta}\bigr)={\tilde  \gamma}^{\alpha \sigma}{\tilde  \gamma}^{\beta \kappa}\bigl({\tilde \mathscr D}_E{\tilde \gamma}_{\sigma \kappa}\bigr)\bigl({\tilde \mathscr D}_P{\tilde \gamma}_{\alpha \beta}\bigr)+{\tilde \mathscr D}_E\bigl({\tilde \gamma}^{\alpha \beta}{\tilde \mathscr D}_P{\tilde \gamma}_{\alpha \beta}\bigr),
\]
we add together the above contributions (from  $\check R_{AC}$ and 
$\check R_{\alpha \beta}$) and get
\begin{eqnarray}
 &&G^{\tilde A\tilde C}N^E_{\tilde A}N^B_{\tilde C}
{\tilde \mathscr D}_E\bigl({\tilde  \gamma}^{\mu \nu}
{\tilde \mathscr D}_B{\tilde \gamma}_{\mu \nu}\bigr)
+\frac14G^{ES}N^P_SN^R_E
\bigl({\tilde \gamma}^{\alpha \beta}{\tilde \mathscr D}_P{\tilde \gamma}_{\nu \beta}\bigr)\bigl({\tilde \gamma}^{\nu \sigma}{\tilde \mathscr D}_R{\tilde \gamma}_{\alpha \sigma}\bigr)\nonumber\\
&&+\frac14G^{ES}N^P_SN^R_E\bigl({\tilde \gamma}^{\alpha \beta}{\tilde \mathscr D}_P{\tilde \gamma}_{\alpha \beta}\bigr)\bigl({\tilde \gamma}^{\mu \nu}{\tilde \mathscr D}_R{\tilde \gamma}_{\mu \nu}\bigr).
\label{term_2r_P}
\end{eqnarray}
Using (\ref{Ricci_2}), (\ref{Ricci_3}), together with (\ref{term_1r_P}) and (\ref{term_2r_P}), in (\ref{scalcurv}),
we obtain the following representation for the scalar curvature: \begin{eqnarray}
 &&R_{\mathcal P}=G^{A' C'}N^S_{A'}N^C_{C'}N^E_M\,{}^{\rm H}R_{SEC}^{\;\;\;\;\;\;\;\;M}+{\tilde \gamma }^{\alpha \beta}{\tilde R}_{\alpha \beta}
\nonumber\\
&&+\frac14\bigl(G^{ES}N^F_SN^B_E\bigr)\,\bigl(G^{MQ}N^P_MN^A_Q\bigr)\,{\tilde \gamma}_{\mu \nu}\,{\tilde \mathcal F}^{\mu}_{PF}{\tilde \mathcal F}^{\nu}_{AB}+G^{A' C'}N^{E}_{A'}\bigl({}^{\rm H}{\tilde\nabla}_{E}({\tilde \gamma}^{\mu \nu}H_{C'}{\tilde \gamma}_{\mu \nu})\bigr)
\nonumber\\
&&+\frac14G^{ES}\bigl({\tilde \gamma}^{\alpha \beta}H_S{\tilde \gamma}_{\nu \beta}\bigr)\bigl({\tilde \gamma}^{\nu \sigma}H_E{\tilde \gamma}_{\alpha \sigma}\bigr)+\frac14G^{ES}\bigl({\tilde \gamma}^{\alpha \beta}H_S{\tilde \gamma}_{\alpha \beta}\bigr)\bigl({\tilde \gamma}^{\mu \nu}H_E{\tilde \gamma}_{\mu \nu}\bigr).
\label{scalcurv_1}
\end{eqnarray}
Here we have used the definition
\[
{}^{\rm H} {\tilde\nabla}_{E}f_C\equiv{\tilde \mathscr D}_E\,f_C-{}^{\rm H}{\Gamma}^M_{CE}f_M.
\]
Notice  that  $R_{\mathcal P}$ is independent of the point in the fiber where it is evaluated.   
It follows from the invariance of 
the original Riemannian metric on $\cal P$ 
under the action of the group $\cal G$.
 So, in (\ref{scalcurv_1}) one can omit  the tilde-marks  placed  over the letters.

\section{The geometrical representation of $\tilde J$}
By comparing the expression for $\tilde J$ given by (\ref{jacobintgrand}) and (\ref{scalcurv_1}), it can be found that $
R_{\mathcal P}$ has the following representation:
\begin{equation}
 R_{\mathcal P}={}^{\rm H}R+ R_{\mathcal G}+ \frac14{\mathcal F}^2+{\tilde J}+\frac14G^{ES}N^A_SN^B_E{\gamma}^{\alpha '\beta '}{\gamma}^{ \nu '\mu '}({\mathscr D}_A {\gamma}_{\nu '\beta '})({\mathscr D}_B {\gamma}_{\mu '\alpha ' }),
\label{scalcurv_2}
\end{equation}
with the evident symbolical notations for 
\begin{eqnarray*}
&&{}^{\rm H}R\equiv G^{A' C'}N^S_{A'}N^C_{C'}N^E_M\,{}^{\rm H}R_{SEC}^{\;\;\;\;\;\;\;\;M},\nonumber\\
&&{\mathcal F}^2\equiv
\bigl(G^{ES}N^F_SN^B_E\bigr)\,\bigl(G^{MQ}N^P_MN^A_Q\bigr)\,{ \gamma}_{\mu \nu}\,{ \mathcal F}^{\mu}_{PF}{ \mathcal F}^{\nu}_{AB},
\end{eqnarray*}
and for  the scalar curvature of the orbit
\[
R_{\mathrm {\cal G}}\equiv\frac12{\gamma}^{\mu\nu} c^{\sigma}_{\mu \alpha} c^{\alpha}_{\nu\sigma}+
\frac14 {\gamma}_{\mu\sigma}{\gamma}^{\alpha\beta}{\gamma}^{\epsilon\nu}
c^{\mu}_{\epsilon \alpha}c^{\sigma}_{\nu \beta}.\]
The last term of (\ref{scalcurv_2}), as it will be shown, is related to the second fundamental form of the orbit.

It follows from the fact that every orbit of the group action
can be locally  viewed as  a submanifold in the  manifold $\cal P$. 
In this case the   second fundamental form of the orbit may be defined as follows:
\[
j^C_{\alpha \beta}(Q)={\Pi}^C_D(Q)\bigl({\nabla}_{K_{\alpha}}K_{\beta})^D(Q),
\]
where by ${\nabla}_A$ we denote the covariant derivative determined by means of the Levy--Civita connection of the manifold $\cal P$ with the  Riemannian metric $G_{AB}(Q)$.
 
We must  project the second fundamental form $j^C_{\alpha \beta}(Q)$ onto the direction which is parallel to the orbit space. 
In order to find this projection we should calculate the following expression:
\begin{equation}
{\tilde G}^{AB}{\tilde G}\left({\Pi}^C_D(Q)\bigl({\nabla}_{K_{\alpha}}K_{\beta}\bigr)^D\frac{\partial}{\partial Q^C},\frac{\partial}{\partial Q^{\ast}{}^A}\right),
\label{secondform}
\end{equation}
where  $\tilde G$ is  the metric (\ref{8}) of the manifold $\cal P$,
and where  before performing the calculation,
the  variables $Q^A$ in
\[
 {\Pi}^C_A(Q)\bigl({\nabla}_{K_{\alpha}}K_{\beta}\bigr)^A(Q)
\frac{\partial}{\partial Q^C}
=\frac12{\Pi}^C_A(Q)\bigl[{\nabla}_{K_{\alpha}}K_{\beta}+{\nabla}_{K_{\beta}}K_{\alpha}\bigr]^A(Q)
\frac{\partial}{\partial Q^C}
\]
must be replaced for $(Q^{\ast}{}^A, a^{\alpha})$.

As a result of the calculation we get
\[
j^B_{\alpha\beta}(Q^{\ast},a)=
\frac12{\rho}^{{\alpha}'}_{\alpha}(a){\rho}^{{\beta}'}_{\beta}(a) N^B_E(Q^{\ast})\bigl({\nabla}_{K_{{\alpha}'}}K_{{\beta}'}+{\nabla}_{K_{{\beta}'}}K_{{\alpha}'}\bigr)^E(Q^{\ast}).
\]
Moreover, it can be shown that
\[
j^B_{\alpha\beta}(Q^{\ast},a)=
-\frac12{\rho}^{\mu}_{\alpha}(a){\rho}^{\nu}_{\beta}(a)\,G^{PS}(Q^{\ast})N^B_P(Q^{\ast})N^E_S(Q^{\ast})
\,\bigl({\mathcal D}_{E}{\gamma}_{\mu\nu}\bigr)(Q^{\ast}).
\]
After restriction of the obtained expression 
to the surface $\Sigma$ by setting $a=e$, 
where $e$ is the unity element of the group $\cal G$,
we come to the following expression for the second fundamental form: 
\[
j^B_{\alpha\beta}(Q^{\ast})=
-\frac12G^{PS}N^B_PN^E_S
\,\bigl({\mathcal D}_{E}{\gamma}_{\alpha \beta}\bigr)(Q^{\ast}).
\]
Using this expression, one can show that the last term of (\ref{scalcurv_2}) is the ``square'' of the fundamental form of the orbit:
\[ 
||j||^2=G^{\rm H}_{AB} \,{\gamma}^{\alpha
\mu}\,{\gamma}^{\beta\nu}\,j^A_{\alpha\beta}\,j^B_{\mu\nu}.
\]

Thus, 
the integrand $\tilde J$ is given by 
\begin{equation}
 {\tilde J}=R_{\mathcal P}-{}^{\rm H}R- R_{\mathcal G}- 
\frac14{\mathcal F}^2-||j||^2.
\label{tilde_J}
\end{equation}

Now, we may rewrite the integral relation (\ref{intgrrelation_1}) in the following form:
\[
{{\gamma}(Q^*_b)}^{-1/4}{{\gamma}(Q^*_a)}^{-1/4}G_{\cal M}(Q^*_b,t_b;Q^*_a,t_a)=
\int_{\cal G}{G}_{\cal P}(p_b\theta ,t_b;
p_a,t_a)d\mu (\theta ),
\]
where
\[
G_{\cal M}(Q^*_b,t_b;Q^*_a,t_a)=\int
d\mu ^{{\xi}_{\Sigma}}
\exp 
\Bigl\{
\int_{t_a}^{t_b}\Bigl[\frac 1{\mu
^2\kappa m}
\tilde V({\xi}_{\Sigma}(u))
-\frac18{\mu^2\kappa m}\,
{\tilde J}({\xi}_{\Sigma}(u))\Bigr ]du
\Bigr\}.
\]
The semigroup determined by the Green's function $G_{\cal M}$ acts in the Hilbert space with the scalar product 
$({\psi}_1,{\psi}_2)=\int_{\Sigma} {\psi}_1(Q^*){\psi}_2(Q^*)\,dv_{\cal M}(Q^*)$.
 The measure $dv_{\cal M}$ is given by
 $dv_{\cal M}(Q^*)=
{\det}^{1/2}\Bigl((P_{\perp})^D_A \;G^{\rm H}_{DC}\,(P_{\perp})^C_B\Bigr)
d{Q^*}^1\wedge\ldots\wedge d{Q^*}^{N_{\cal P}}.
$

If it were possible to find invariant coordinates $x^i$ such that
${\chi}^{\alpha}(Q^*(x^i))\equiv 0$, the 
 measure $dv_{\cal M}$ of the previouse scalar product could be transformed into the volume measure ${\det}^{1/2} h_{ij}\,\,dx^1\cdot\ldots \cdot dx^{N_{\cal M}}$ for  the Riemannian metric  $h_{ij}={Q^*}^A_i(x)G^{\rm H}_{AB}({Q^*}(x)){Q^*}^B_j(x)$  defined  on the orbit space $\cal M$.

The Green's function $G_{\cal M}$ satisfies the forward Kolmogorov equation 
with the operator
\begin{eqnarray}
 &&{\hat H}_{\kappa}=\frac{\hbar\kappa}{2m} \biggl\{
 G^{CD}N^A_CN^B_D\frac{{\partial}^2}{\partial
 Q^{*}{}^A\partial
 Q^{*}{}^B}-G^{CD}N^E_CN^M_D\,{}^H{\Gamma}^A_
 {EM}\frac{\partial}{\partial
 Q^*{}^A}
\biggr\} 
\nonumber\\
&&\;\;\;\;\;\;\;\;\;\;\;\;\;\;\;\;\;\;\;
-\frac{\hbar\kappa}{8m}{\tilde J}+\frac{1}{\hbar\kappa}
{\tilde V},
 \label{op3}
 \end{eqnarray} 
where  $\tilde J$ is given by (\ref{tilde_J}).
The Hamilton operator $\hat H$ of the corresponding Schr\"odinger equation can be obtained from (\ref{op3})
as follows:  $\hat H=-\frac{\hbar}{\kappa}{\hat H}_{\kappa}|_{\kappa =i}$.

\section{Conclusion} 
In the paper, it has been shown  that   the exponential of the path integral reduction Jacobian can be written in  the form of  the difference between  the scalar curvature of the original manifold and the following terms: the scalar curvature of the orbit, the scalar curvature of the reduced manifold, the square of the second fundamental form of the orbit, and the one fourth  of the square of the  curvature of the connection defined on the principal fibre bundle.  

In  many important cases the local description of the reduced motion is only possible by making use of  dependent coordinates. 
This is a typical situation  which  one meets with in gauge  theories.
It would be very useful to find an appropriate generalization of the obtained formula (\ref{tilde_J}) in these cases. 

Besides, the formulae of this kind are necessary  for consideration of the renormalization corrections in case of the rigorous definition of the path integral measure defined on the space of  gauge connections, where  the regularization of the original (weak) metric converts it into the (strong) Riemannian metric \cite{Asorey}.

In the paper the geometrical representation of the Jacobian  has been found, in fact, for  the case of the local reduction since consideration of \cite{Storchak_2} has been done for the trivial principal fibre bundle. 
An interesting problem would be to extend the obtained result  onto the case of the global path integral reduction in which the topological questions 
would be expected to play an important role.

\section*{Acknowledgment}
The author is 
grateful to  Yu. M. Zinoviev, A. V. Razumov and V. O. Soloviev
for very stimulating discussions.

\end{document}